\documentclass{elsart1p}
 \usepackage{graphics}
\usepackage{graphicx}
\usepackage{amssymb}

\newcommand{\lambdabar}{{\hbox{$\lambda$\kern-1.ex\raise+0.45ex\hbox{--}}}}

\begin{document}
\begin{frontmatter}
% use the corauthref command within \author for corresponding author
% footnotes;
\title{
{\small\mbox{}\hfill DESY 09-001}\\[1.5ex]
Prospects for the direct detection of the cosmic neutrino background\thanksref{label1}}
 \thanks[label1]{Invited talk presented at PANIC 2008, 9-14 November 2008, Eilat, Israel}
%Prospects for the direct detection of the cosmic neutrino background}
 \author{Andreas Ringwald}
% \ead{andreas.ringwald@desy.de}
% \ead[url]{www.desy.de/\~{}ringwald}
 \address{Deutsches Elektronen-Synchrotron DESY, Notkestrasse 85, D-22607 Hamburg, Germany}
%
%\title{Prospects for the direct detection of the cosmic neutrino background}
\author{}
\address{}
\begin{abstract}
The existence of a cosmic neutrino background -- the analogue of the 
cosmic microwave background -- is a fundamental prediction of standard big 
bang cosmology. Up to now, the observational evidence for its existence 
is rather indirect and rests entirely on cosmological observations of, e.g.,   
the light elemental abundances, the anisotropies in the cosmic microwave
background, and the large scale distribution of matter. Here, we review
more direct, weak interaction based detection techniques for the cosmic 
neutrino background in the present epoch and in our local neighbourhood.      
We show that, with current technology, all proposals are still off by some orders of magnitude in
sensitivity to lead to a guaranteed detection of the relic neutrinos. 
The most promising laboratory search, based on neutrino capture 
on beta decaying nuclei, may be done in  
future experiments designed to measure the neutrino mass through decay kinematics. 
\end{abstract}
%
%\begin{keyword}
%\PACS
%\end{keyword}
\end{frontmatter}
%
% main text
\section{Introduction}
\label{intro}

Over the past decade, we have witnessed extraordinary progress in observational
cosmology. In fact, from the interpretation of cosmological data a quite precise
knowledge of the energy (mass) budget of the universe emerged. Along with the
heavily exploited cosmic microwave background (CMB), standard big bang theory predicts the 
existence of a cosmic neutrino background (CNB), comprised of relic neutrinos from the
epoch of decoupling of weak interactions about one second after the big bang. 
Presently, the inference about or evidence for the existence of the CNB rests solely on cosmological
measurements, in particular the light elemental abundances and their interpretation in
terms of big bang nucleosynthesis (BBN), the spectrum of CMB anisotropies, and the 
large scale matter power spectrum (for a review, see Ref.~\cite{Hannestad:2006zg}). 
All these measurements, however, probe the
presence of relic neutrinos only at early stages in the cosmological evolution and, moreover,
in a rather indirect way. Here, we will concentrate on more direct, weak interaction
based detection possibilities of the CNB, sensitive in particular to the CNB 
in our local neighborhood and in the present epoch.

\section{Phase space distribution of relic neutrinos}
\label{clustering}

The design of a direct, weak interaction based detection experiment needs a 
precise knowledge of the phase space distribution of the relic neutrinos. 
Its average over large scales has been determined when the primordial 
plasma had a temperature of about one MeV: it is given by the homogeneous and 
isotropic relativistic Fermi-Dirac distribution, $f_0=1/(1+\exp (p/T_{\nu ,0}))$,
where $p$ is the modulus of the comoving three-momentum $\mathbf p$ and 
$T_{\nu ,0}=(4/11)^{1/3} T_{\gamma ,0}=1.95$~K is today's CNB 
temperature, in terms of the CMB temperature $T_{\gamma ,0}=2.73$~K.
Correspondingly, the large scale properties of the CNB are tightly 
related to the properties of the well-measured CMB and are therefore
to be considered as fundamental predictions of standard big bang cosmology. 
Their present number density,
\begin{equation}
\label{bar_n}
\underbrace{{\bar n}_{\nu , 0} = {\bar n}_{\bar \nu ,0}}_{\rm CNB} 
= \frac{3}{22} \underbrace{{\bar n}_{\gamma ,0}}_{\rm CMB} = 56\ {\rm cm}^{-3}, 
\end{equation}  
when summed over all netrino types $i=1,2,3$, is large and comparable to the one
of the CMB. Their present average three-momentum, on the other hand, is very 
small,
\begin{equation}
\label{bar_p}
\underbrace{{\bar p}_{\nu ,0} = {\bar p}_{\bar \nu ,0}}_{\rm CNB} =
3\, (4/11)^{1/3}\, \underbrace{T_{\gamma ,0}}_{\rm CMB} = 5\times 10^{-4}\ {\rm eV} . 
\end{equation}
Correspondingly, at least two of the relic neutrino mass eigenstates are non-relativistic
today ($m_{\nu_i}\gg {\bar p}_{\nu_i,0}$), independently of whether neutrino masses have 
a normal hierarchical or inverted hierarchical pattern. These neutrinos are 
subject to gravitational clustering into gravitational potential wells due to existing
cold dark matter (CDM) and baryonic structures, causing the local neutrino number density
to be enhanced relative to the standard value~(\ref{bar_n})  
and the momentum distribution to deviate from the one following
from the Fermi-Dirac distribution. 

\begin{figure}
\centerline{
\includegraphics[width=0.5\textwidth]{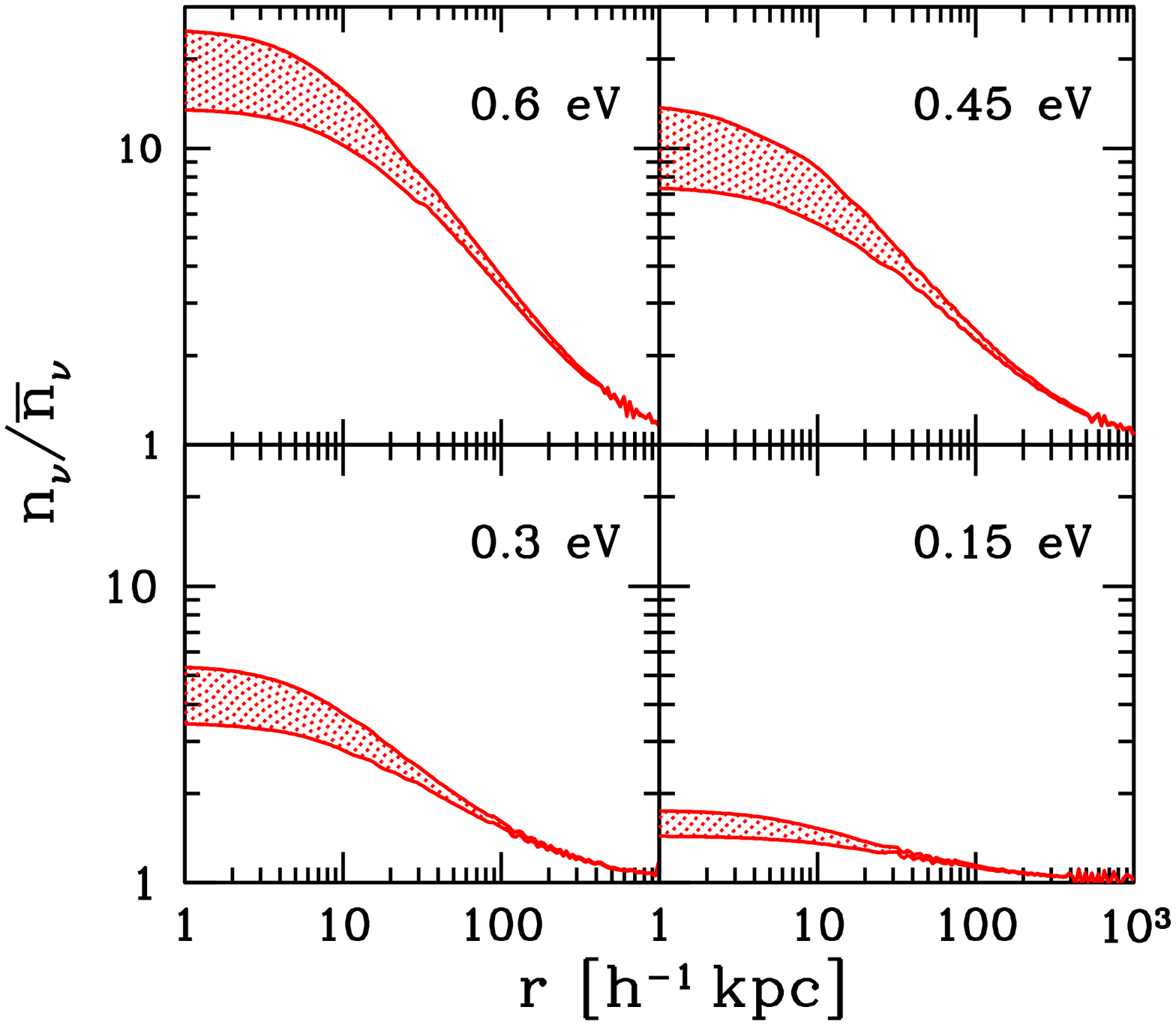}
\includegraphics[width=0.5\textwidth]{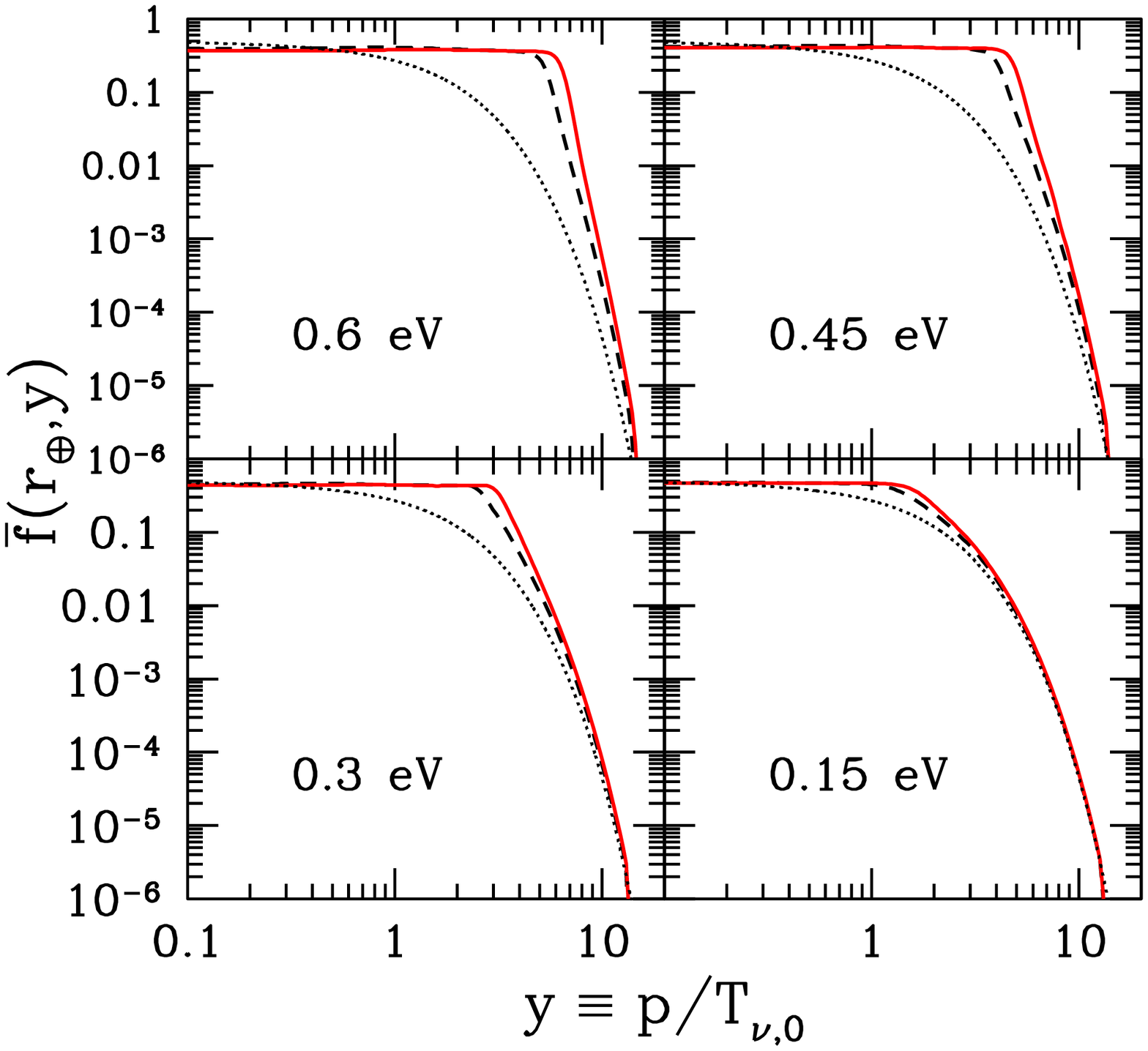}
}
\caption{Characteristics of the CNB in the Milky Way, obtained from solutions
of the collisionless Boltzmann equation for the neutrino phase space distribution in the background
of the gravitational potential of a NFW CDM halo (NFWhalo) corresponding to the Milky 
Way and of the present day Milky Way mass distribution 
(MWnow)~\cite{Ringwald:2004np}. The results are displayed
for different values
of the neutrino mass, ranging from $0.15$ to $0.6$~eV.  
 {\em Left panels:} Neutrino number density profiles, normalized to their cosmological 
mean, as a function of the distance to the galactic center. The top (bottom) curves correspond
to the MWnow (NFWhalo) gravitational potential. 
{\em Right panels:} Momentum distribution of the relic neutrinos in the local neighborhood
of the Earth, obtained from the MWnow (solid) and NFWhalo (dashed) potential, approaching
for large momenta the relativistic Fermi-Dirac distribution (dotted).
}\label{overdens_mw}
\end{figure}

A comprehensive and exhaustive study of gravitational clustering of relic neutrinos~\cite{Ringwald:2004np} 
has revealed that, within the range of possible neutrino 
masses\footnote{The present cosmological limit on the sum of the neutrino masses is in the  
$\sum_i m_{\nu_i}<0.5-0.6$~eV range for a $\Lambda$CDM model, but can be relaxed by a large factor if 
more parameters are included~\cite{Hannestad:2006zg}.}, one can expect an overdensity of $1\div 20$ over the 
mean values in our position in the Milky Way, i.e. at a distance of about $r_\oplus = 8$~kpc from the galactic 
center (c.f. Fig.~\ref{overdens_mw} (left)). The present day momentum
distribution near the Earth is found to be almost isotropic, with mean radial velocity 
$\langle v_r\rangle \approx 0$ and second moments that satisfy approximately the 
relation $2\langle v_r^2\rangle \approx \langle v_T^2\rangle$. The coarse-grained
phase space densities $\bar f(r_\oplus , p)$ in Fig.~\ref{overdens_mw} (right) are flat 
at low momenta, with a common value of nearly 1/2, have a turning point at about the escape
momenta corresponding to the gravitational potential, and quickly approach the Fermi-Dirac distribution
for larger momenta.    

\section{Direct CNB detection techniques and their prospects}
\label{prospects}

In this section we will contentrate on detection techniques of the  
CNB which are based on weak interaction scattering processes involving the 
relic neutrinos as beam or as target. 

\subsection{Detection via mechanical force due to relic neutrino elastic scattering off target}
\label{force_based_detection}

\begin{figure}
\centerline{
\includegraphics[bbllx=80pt,bblly=337pt,bburx=586pt,bbury=662pt,width=0.7\textwidth,clip=]{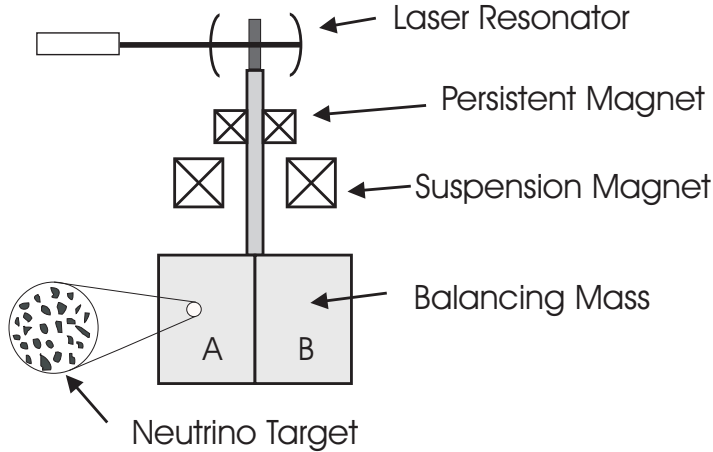}
}
\caption{Cavendish-type torsion balance for relic neutrino detection~\cite{Hagmann:1998nz}. 
}\label{torsion_balance}
\end{figure}

The Earth is moving through the almost isotropic (cf. last section)
relic neutrino background. Coherent scattering of the corresponding
relic neutrino flux off target matter in a terrestrial detector will
lead to a mechanical force~\cite{Shvartsman:1982sn,Smith:1983jj} which may be detected in Cavendish-type 
torsion balances (cf. Fig.~\ref{torsion_balance}) by searching for an annual modulation of the signal. 
In fact, a terrestrial detector of mass density $\rho_{\rm t}$ and 
linear size $r_{\rm t}<\lambdabar$, where 
\begin{equation} 
\lambdabar = 1/\langle p\rangle = 0.12\ {\rm ~cm}/\langle p/T_{\nu ,0}\rangle ,
\end{equation}   
is the de Broglie wavelength of the relic neutrinos, 
will experience a neutrino wind induced acceleration due to 
coherent scattering~\cite{Shvartsman:1982sn,Smith:1983jj,Ferreras:1995wf,Hagmann:1998nz,Duda:2001hd}, 
\begin{eqnarray} 
a_{\rm t} &\simeq & \sum_{\nu,\bar\nu}\  
\underbrace{n_{\nu}\,v_{\rm rel}}_{\rm flux}\ 
\frac{4\pi}{3}\, N_A^2\, \rho_{\rm t}\,  r_{\rm t}^3 
\  
\sigma_{\nu N}\,  
\underbrace{2\,m_\nu\,v_{\rm rel}}_{\rm mom.\, transfer}
\label{acceler} 
\\[-.ex] \nonumber 
&\simeq & 
{2\times 10^{-28}}  {\ \frac{\rm cm}{{\rm s}^{2}}}
\left( \frac{n_\nu}{\bar n_\nu}\right)  
\left( \frac{10^{-3}\,c}{v_{\rm rel}}\right) 
\left( \frac{\rho_{\rm t}}{{\rm g/cm^3}}\right) 
\left( \frac{r_{\rm t}}{\lambdabar}\right)^3 ,  
\end{eqnarray}
where $N_A$ is Avogadro's number, $\sigma_{\nu N}\simeq 
G_F^2 m_\nu^2/\pi$ is the elastic neutrino nucleon cross section,
and $v_{\rm rel} = \langle |{\mathbf v}-{\mathbf v}_\oplus|\rangle$  
is the mean velocity of the relic neutrinos in the rest 
system of the detector. Here $v_\oplus \simeq 7.7\times 10^{-4}\,c$ 
denotes the velocity of the Earth through the Milky Way. 
The acceleration~(\ref{acceler}) can be improved further by using 
foam-like~\cite{Shvartsman:1982sn} or laminated~\cite{Smith:1983jj} materials, 
or by embedding grains of 
size $\lambdabar$ (with spacing $\sim\lambdabar$) randomly in a low density 
host material~\cite{Smith:2003sy}
(cf. Fig.~\ref{torsion_balance}).
In this way one may exploit a target size much   
larger than $\lambdabar$, while still avoiding destructive interference. 
For Majorana type neutrinos, the acceleration has a further suppression factor    
$(v_{\rm rel}/c)^2\simeq 10^{-6}$ for an unpolarized and 
$v_{\rm rel}/c\simeq 10^{-3}$ for a polarized target~\cite{Hagmann:1998nz}, respectively. 

To infer the prospects of this CNB detection method we note that
presently Cavendish-type torsion balances routinely reach an acceleration sensitivity   
of $10^{-13}$~cm/s$^2$~\cite{Adelberger:2009zz}. Improvements obtainable with current technology 
may improve this down to $10^{-23}$~cm/s$^2$~\cite{Hagmann:1998nz} (cf. Fig.~\ref{torsion_balance}). 
Correspondingly, at present this method falls short in sensitivity by at least three orders
of magnitude. But it can still be envisaged in the not-so-distant future, say, 
within thirty to forty years. For the (most likely) case that the neutrinos are of Majorana type, 
however, the opportunities of CNB detection via this method seem to be very slim. 

\subsection{Detection via relic neutrino capture on radioactive nuclei}
\label{capture}

\begin{figure}[t]
\centerline{
\includegraphics[bbllx=4pt,bblly=2pt,bburx=386pt,bbury=210pt,width=0.8\textwidth,clip=]{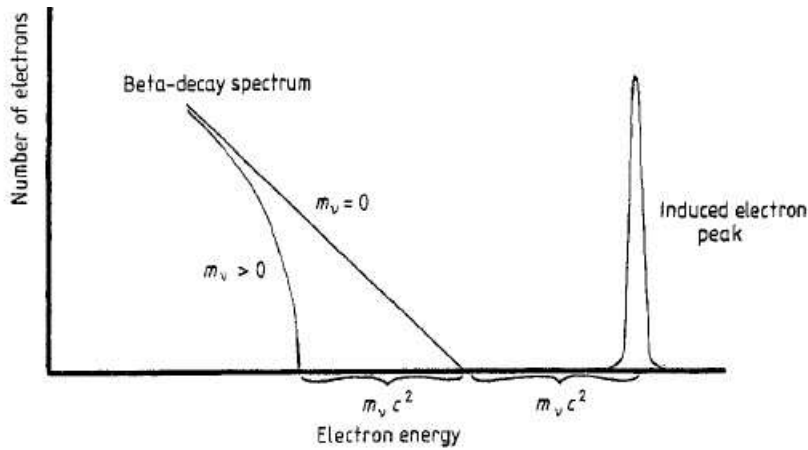}
}
\caption{Idealized electron spectrum for tritium beta decay plus relic neutrino capture from Ref.~\cite{Irvine:1983nr}. 
}\label{beta_decay_spectrum_idealized}
\end{figure}

A very promising technique for direct detection of the 
CNB may be realized in beta decay experiments initially designed to
measure the neutrino mass\footnote{For an interesting proposal to detect relic neutrinos via Pauli blocking effects   
near thresholds for atomic neutrino pair mission enhanced by laser irradiation, see Ref.~\cite{Takahashi:2007ec}.}   
through the kinematics of the 
decay~\cite{Irvine:1983nr,Cocco:2007za,Lazauskas:2007da,Blennow:2008fh}. 
It takes advantage of the fact that the rate for relic neutrino 
capture on beta decaying nuclei, e.g. on tritium,   
$\nu_i +\, ^3\mathrm{H} \rightarrow
e +\, ^3\mathrm{He}$, 
converges to a finite value for $v_\nu\to 0$, e.g.
\begin{equation}
N_{i,\,\mathrm{CNB}} \simeq  6.5\ \mathrm{yr}^{-1} 
\left( 100\,\mathrm{g}\ ^3\mathrm{H}\right)^{-1}
|U_{ei}|^2\  
\frac{n_{\nu_i}}{\bar{n}_{\nu_i}} ,
\end{equation}
where $|U_{ei}|^2$ is the electron 
neutrino content of $\nu_i$. 
Correspondingly, for a sufficient amount of beta decaying target material,
the capture rate is reasonably large. Furthermore, because of the non-zero neutrino mass,
this process has also a unique signature provided by monoenergetic 
electrons with kinetic energy $Q_\beta + m_{\nu_i}$, where 
$Q_\beta$ is the energy release for beta decay with $m_{\nu_i}=0$, 
e.g. $Q_\beta =18.6$~keV for tritium. This is illustrated in Fig.~\ref{beta_decay_spectrum_idealized} 
for an idealized situation where $|U_{ei}|^2=\delta_{ei}$ and where the detector energy resolution 
as well as experimental background has been neglected.      

What are the prospects of this technique? 
At presently developed beta decay experiments, the relatively small amount of target material 
leads to a tiny total rate of relic neutrio capture. Furthermore, the energy resolution is still too large, $\Delta E \gtrsim m_{\nu_i}$,  and 
the background rate is too high to expect a discovery of the CNB in the very near future.  
For example, the tritium beta decay experiment KATRIN~\cite{Osipowicz:2001sq} will exploit in its first phase starting 
in 2012 effectively 
$\sim 5\times 10^{18}$ T$_2$ molecules, corresponding to $\sim 5\times 10^{-5}$~g tritium~\cite{Weinheimer:2008}, 
is foreseen to have an energy resolution $\Delta E\sim 1$~eV and a background rate of about 10~mHz. 
Assuming a neutrino mass $m_{\nu_i}= 0.6$~eV and taking gravitational clustering into account, 
a 5~$\sigma$ evidence for the CNB may be achieved in one year if it were possible, in the second phase of KATRIN, which aims at
$\Delta E\sim 0.2$~eV and a background rate of 1~mHz, to  increase the effective mass of tritium to $\sim 35$~g~\cite{Messina:2008}. 
The $^{187}$Re beta decay experiment MARE~\cite{Andreotti:2007eq}, on the other hand, foresees to have, in the year  
$\sim 2011$, $10^5$ micro-calorimeters of $1-5$~mg, which is still 4-6 orders of magnitude below the 
mass needed for a reasonable capture rate~\cite{Messina:2008}. But the micro-calorimetric approach seems to be
scalable quite easily.   

\subsection{Detection via absorption features in extremely energetic cosmic neutrino spectra}

\begin{figure}[t]
\centerline{
\includegraphics[bbllx=21pt,bblly=218pt,bburx=565pt,bbury=673pt,width=0.6\textwidth,clip=]{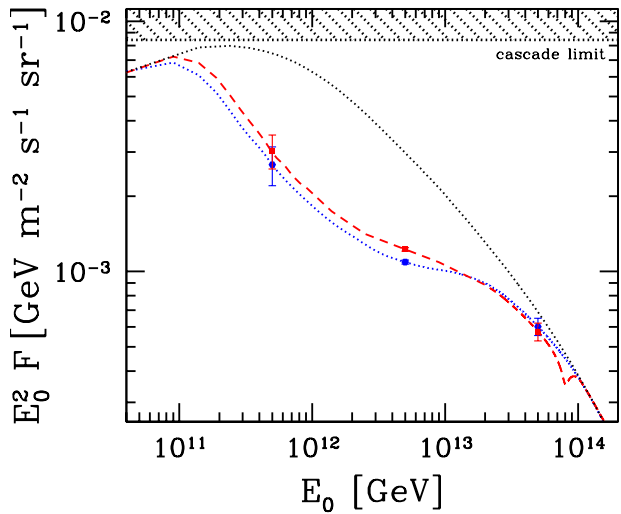}
}
\caption{EHEC$\nu$ spectrum from decaying superheavy ($m_X=10^{16}$~GeV) particles 
radiated off cosmic strings 
(cf. Ref.~\cite{Ringwald:2006ks}), taking into account relic neutrino absorption for a hierarchical neutrino 
spectrum with $m_{\nu_1}=10^{-5}$~eV, $m_{\nu_2}=8.3\times 10^{-3}$~eV, $m_{\nu_3}=5.2\times 10^{-2}$~eV (dashed line).
The error bars correspond to the combined projected sensitivity of the ANITA~\cite{Gorham:2008dv} and LOFAR~\cite{Scholten:2006ja} 
EHEC$\nu$ detectors.  
}\label{ehecnu_spectrum}
\end{figure}

An appealing opportunity to catch a glimpse of the CNB in the present epoch emerges from the possible existence of 
extremely high-energy cosmic neutrinos (EHEC$\nu$'s), originating e.g. from decaying superheavy particles which are 
radiated off topological defects generated at phase transitions in the very early universe (cf. Fig.~\ref{ehecnu_spectrum}). 
EHEC$\nu$s can annihilate with relic anti-neutrinos (and vice versa) into $Z$ bosons, $\nu\bar\nu\rightarrow Z$, 
if their energies coincide with the respective resonance energies,
\begin{equation}
\label{Eres0}
E_{0,i}^{\rm res}=\frac{m^2_Z}{2m_{\nu_{0,i}}}=4.2\times 10^{12}\,\,\left(\frac{\rm eV}{m_{\nu_i}}\right) {\rm GeV},
\end{equation}
where $m_Z$ denotes the $Z$ mass.
An exceptional loss of transparency of the CNB for cosmic neutrinos results from the fact that the 
corresponding annihilation cross-section on resonance is enhanced by several orders of magnitude with respect to non-resonant scattering. 
As a consequence, the diffuse EHEC$\nu$ flux arriving at Earth is expected to exhibit absorption dips~\cite{Weiler:1982qy,Eberle:2004ua,Barenboim:2004di,D'Olivo:2005uh,Ringwald:2006ks,Scholten:2008xg} 
whose locations 
in the spectrum are determined by the respective resonance energies of the annihilation processes. Provided that 
the dips can be resolved on Earth, they could produce the most direct evidence for the existence of the CNB so far. 
Promisingly, EHEC$\nu$ detectors such as ANITA~\cite{Gorham:2008dv}, LOFAR~\cite{Scholten:2006ja} and 
SKA~\cite{James:2008ff}, which will operate within the next decade, 
have a very good sensitivity 
in the relevant energy range. In fact, provided that the actual EHEC$\nu$ flux is close to the current bounds, relic neutrino absorption spectroscopy 
can become a realistic possibility in the near future~(cf. Fig.~\ref{ehecnu_spectrum}). 

%
%\section{Conclusions}
%\label{conclusions}
%

% \appendix
%
% \section{}
% \label{}
%

%

\begin{thebibliography}{00}
% notes:
% \bibitem{label2}
% If there is a note, it should come last:
% \bibitem{label3} \note
% \end{subbibitems}
%

%\cite{Hannestad:2006zg}
\bibitem{Hannestad:2006zg}
  S.~Hannestad,
  %``Primordial Neutrinos,''
  Ann.\ Rev.\ Nucl.\ Part.\ Sci.\  {\bf 56} (2006) 137
  [arXiv:hep-ph/0602058].
  %%CITATION = ARNUA,56,137;%%

%\cite{Ringwald:2004np}
\bibitem{Ringwald:2004np}
  A.~Ringwald and Y.~Y.~Y.~Wong,
  %``Gravitational clustering of relic neutrinos and implications for their
  %detection,''
  JCAP {\bf 0412} (2004) 005
  [arXiv:hep-ph/0408241].
  %%CITATION = JCAPA,0412,005;%%

%\cite{Shvartsman:1982sn}
\bibitem{Shvartsman:1982sn}
  B.~F.~Shvartsman {\em et al.}, 
%, V.~B.~Braginsky, S.~S.~Gershtein, Y.~B.~Zeldovich and M.~Y.~Khlopov,
  %``Possibility Of Detecting Relict Massive Neutrinos,''
  JETP Lett.\  {\bf 36} (1982) 277
  [Pisma Zh.\ Eksp.\ Teor.\ Fiz.\  {\bf 36} (1982) 224].
  %%CITATION = ZFPRA,36,224;%%

%\cite{Smith:1983jj}
\bibitem{Smith:1983jj}
  P.~F.~Smith and J.~D.~Lewin,
  %``Coherent Interaction Of Galactic Neutrinos With Material Targets,''
  Phys.\ Lett.\  B {\bf 127} (1983) 185.
  %%CITATION = PHLTA,B127,185;%%

%\cite{Ferreras:1995wf}
\bibitem{Ferreras:1995wf}
  I.~Ferreras and I.~Wasserman,
  %``Feasibility of observing mechanical effects of cosmological neutrinos,''
  Phys.\ Rev.\  D {\bf 52} (1995) 5459.
  %%CITATION = PHRVA,D52,5459;%%

%\cite{Hagmann:1998nz}
\bibitem{Hagmann:1998nz}
  C.~Hagmann,
  %``A relic neutrino detector,''
  arXiv:astro-ph/9902102;
  %%CITATION = ASTRO-PH/9902102;%%
%
%\cite{Hagmann:1999kf}
%\bibitem{Hagmann:1999kf}
%  C.~Hagmann,
  %``Cosmic neutrinos and their detection,''
  arXiv:astro-ph/9905258.
  %%CITATION = ASTRO-PH/9905258;%%

%\cite{Duda:2001hd}
\bibitem{Duda:2001hd}
  G.~Duda, G.~Gelmini and S.~Nussinov,
  %``Expected signals in relic neutrino detectors,''
  Phys.\ Rev.\  D {\bf 64} (2001) 122001
  [arXiv:hep-ph/0107027].
  %%CITATION = PHRVA,D64,122001;%%

%\cite{Smith:2003sy}
\bibitem{Smith:2003sy}
  P.~F.~Smith,
  %``Direct detection of weakly interacting massive particles using
  %non-cryogenic techniques,''
  Phil.\ Trans.\ Roy.\ Soc.\ Lond.\  A {\bf 361} (2003) 2591.
  %%CITATION = PTRSA,A361,2591;%%

%\cite{Adelberger:2009zz}
\bibitem{Adelberger:2009zz}
  E.~G.~Adelberger {\em et al.}, 
%J.~H.~Gundlach, B.~R.~Heckel, S.~Hoedl and S.~Schlamminger,
  %``Torsion balance experiments: A low-energy frontier of particle physics,''
  Prog.\ Part.\ Nucl.\ Phys.\  {\bf 62} (2009) 102.
  %%CITATION = PPNPD,62,102;%%

%\cite{Takahashi:2007ec}
\bibitem{Takahashi:2007ec}
  T.~Takahashi and M.~Yoshimura,
  %``Effect of relic neutrino on neutrino pair emission from metastable atoms,''
  arXiv:hep-ph/0703019.
  %%CITATION = HEP-PH/0703019;%%

%\cite{Irvine:1983nr}
\bibitem{Irvine:1983nr}
  J.~M.~Irvine and R.~Humphreys,
  %``Neutrino Masses And The Cosmic Neutrino Background,''
  J.\ Phys.\ G {\bf 9} (1983) 847.
  %%CITATION = JPHGB,G9,847;%%

%\cite{Cocco:2007za}
\bibitem{Cocco:2007za}
  A.~G.~Cocco, G.~Mangano and M.~Messina,
  %``Probing low energy neutrino backgrounds with neutrino capture on beta
  %decaying nuclei,''
  JCAP {\bf 0706} (2007) 015
%  [J.\ Phys.\ Conf.\ Ser.\  {\bf 110} (2008) 082014]
  [arXiv:hep-ph/0703075].
  %%CITATION = 00462,110,082014;%%

%\cite{Lazauskas:2007da}
\bibitem{Lazauskas:2007da}
  R.~Lazauskas, P.~Vogel and C.~Volpe,
  %``Charged current cross section for massive cosmological neutrinos impinging
  %on radioactive nuclei,''
  J.\ Phys.\ G {\bf 35} (2008) 025001
  [arXiv:0710.5312 [astro-ph]].
  %%CITATION = JPHGB,G35,025001;%%

%\cite{Blennow:2008fh}
\bibitem{Blennow:2008fh}
  M.~Blennow,
  %``Prospects for cosmic neutrino detection in tritium experiments in the case
  %of hierarchical neutrino masses,''
  Phys.\ Rev.\  D {\bf 77} (2008) 113014
  [arXiv:0803.3762 [astro-ph]].
  %%CITATION = PHRVA,D77,113014;%%

%\cite{Osipowicz:2001sq}
\bibitem{Osipowicz:2001sq}
  A.~Osipowicz {\it et al.}  [KATRIN Collaboration],
  %``KATRIN: A next generation tritium beta decay experiment with sub-eV
  %sensitivity for the electron neutrino mass,''
  arXiv:hep-ex/0109033.
  %%CITATION = HEP-EX/0109033;%%

%\cite{Weinheimer:2008}
\bibitem{Weinheimer:2008}
 C.~Weinheimer, talk presented at Workshop on New Instruments for Neutrino Relics and Mass, 
 8 December 2008, CERN, Geneva/CH, http://indico.cern.ch/confAuthorIndex.py?confId=42884 

%\cite{Messina:2008}
\bibitem{Messina:2008}
 M.~Messina, talk presented at Workshop on New Instruments for Neutrino Relics and Mass, 
 8 December 2008, CERN, Geneva/CH, http://indico.cern.ch/confAuthorIndex.py?confId=42884 

%\cite{Andreotti:2007eq}
\bibitem{Andreotti:2007eq}
  E.~Andreotti {\it et al.},
  %``MARE, Microcalorimeter Arrays for a Rhenium Experiment: A detector
  %overview,''
  Nucl.\ Instrum.\ Meth.\  A {\bf 572} (2007) 208.
  %%CITATION = NUIMA,A572,208;%%

%\cite{Weiler:1982qy}
\bibitem{Weiler:1982qy}
  T.~J.~Weiler,
  %``Resonant Absorption Of Cosmic Ray Neutrinos By The Relic Neutrino
  %Background,''
  Phys.\ Rev.\ Lett.\  {\bf 49} (1982) 234;
  %%CITATION = PRLTA,49,234;%%
%
%\cite{Weiler:1983xx}
%\bibitem{Weiler:1983xx}
%  T.~J.~Weiler,
  %``Big Bang Cosmology, Relic Neutrinos, And Absorption Of Neutrino Cosmic
  %Rays,''
  Astrophys.\ J.\  {\bf 285}, 495 (1984).
  %%CITATION = ASJOA,285,495;%%

%\cite{Eberle:2004ua}
\bibitem{Eberle:2004ua}
  B.~Eberle, A.~Ringwald, L.~Song and T.~J.~Weiler,
  %``Relic neutrino absorption spectroscopy,''
  Phys.\ Rev.\  D {\bf 70} (2004) 023007.
 % [arXiv:hep-ph/0401203].
  %%CITATION = PHRVA,D70,023007;%%

%\cite{Barenboim:2004di}
\bibitem{Barenboim:2004di}
  G.~Barenboim, O.~Mena Requejo and C.~Quigg,
  %``Diagnostic potential of cosmic-neutrino absorption spectroscopy,''
  Phys.\ Rev.\  D {\bf 71} (2005) 083002.
 % [arXiv:hep-ph/0412122].
  %%CITATION = PHRVA,D71,083002;%%

%\cite{D'Olivo:2005uh}
\bibitem{D'Olivo:2005uh}
  J.~C.~D'Olivo, L.~Nellen, S.~Sahu and V.~Van Elewyck,
  %``UHE neutrino damping in a thermal gas of relic neutrinos,''
  Astropart.\ Phys.\  {\bf 25} (2006) 47.
 % [arXiv:astro-ph/0507333].
  %%CITATION = APHYE,25,47;%%

%\cite{Ringwald:2006ks}
\bibitem{Ringwald:2006ks}
  A.~Ringwald and L.~Schrempp,
  %``Probing neutrino dark energy with extremely high-energy cosmic
  %neutrinos,''
  JCAP {\bf 0610}, 012 (2006)
  [arXiv:astro-ph/0606316].
  %%CITATION = JCAPA,0610,012;%%

%\cite{Scholten:2008xg}
\bibitem{Scholten:2008xg}
  O.~Scholten and A.~van Vliet,
  %``Determining neutrino absorption spectra at Ultra-High Energies,''
  JCAP {\bf 0806} (2008) 015
  [arXiv:0801.3342 [astro-ph]].
  %%CITATION = JCAPA,0806,015;%%

%\cite{Gorham:2008dv}
\bibitem{Gorham:2008dv}
  P.~Gorham {\it et al.}  [ANITA collaboration],
  %``The Antarctic Impulsive Transient Antenna Ultra-high Energy Neutrino
  %Detector Design, Performance, and Sensitivity for 2006-2007 Balloon Flight,''
  arXiv:0812.1920 [astro-ph]; 
  %%CITATION = ARXIV:0812.1920;%%
%
%\cite{Gorham:2008yk}
%\bibitem{Gorham:2008yk}
%  P.~Gorham {\it et al.}  [ANITA collaboration],
  %``New Limits on the Ultra-high Energy Cosmic Neutrino Flux from the ANITA
  %Experiment,''
  arXiv:0812.2715 [astro-ph].
  %%CITATION = ARXIV:0812.2715;%%

%\cite{Scholten:2006ja}
\bibitem{Scholten:2006ja}
  O.~Scholten {\em et al.}, 
%, J.~Bacelar, R.~Braun, A.~G.~de Bruyn, H.~Falcke, B.~Stappers and R.~G.~Strom,
  %``Optimal radio window for the detection of ultra-high-energy cosmic rays
  %and neutrinos off the moon,''
  Astropart.\ Phys.\  {\bf 26} (2006) 219.
  % [arXiv:astro-ph/0609179].
  %%CITATION = APHYE,26,219;%%

%\cite{James:2008ff}
\bibitem{James:2008ff}
  C.~W.~James and R.~J.~Protheroe,
  %``The sensitivity of the next generation of lunar Cherenkov observations to
  %UHE neutrinos and cosmic rays,''
  arXiv:0802.3562 [astro-ph].
  %%CITATION = ARXIV:0802.3562;%%

%
\end{thebibliography}
\end{document}